\documentclass[superscriptaddress,10pt,twocolumn]{revtex4-2} 

\usepackage{mathtools}
\usepackage{amsmath}  
\usepackage{amsfonts} 
\usepackage{graphicx} 
\usepackage{xcolor}
\usepackage{bm}

\usepackage[justification=justified,labelfont=bf,format=plain]{caption}
\DeclareCaptionFormat{myformat}{\fontsize{10}{10} \selectfont#1#2#3}
\begin{document}
\title{Holographic model for color superconductivity \\ in d-dimension bulk without confinement phase}

\author{Nguyen Hoang Vu}
    \email[Correspondence email address: ]{vu@jinr.ru}
    \affiliation{The Bogoliubov Laboratory of Theoretical Physics, JINR,141980 Dubna, Moscow region, Russia}
    \affiliation{Institute of Physics, VAST, 10000, Hanoi, Vietnam}


\date{\today} 

\begin{abstract}
We generalize the concept of holography for the color superconductivity (CSC) phase by considering $d$-dimensional Anti de Sitter (AdS) space instead of the traditional 6 dimensions. The corresponding dual field theory is a gauge theory with $SU(N_c)$ symmetry defined in $(d-1)$-dimensions that, despite lacking a confinement phase, retains characteristics consistent with quantum chromodynamics (QCD) CSC. We then use a holographic model based on Einstein-Maxwell gravity and the standard Maxwell interaction in $d$-dimensional AdS space to investigate this phenomenon for the number of colors $N_c\geq 2$ without confinement phase, identifying the dimensions where the model remains valid for $N_c = 2$.
\end{abstract}

\keywords{CSC phase, Einstein-Maxwell gravity, AdS/CFT}

\maketitle

\section{Introduction}
In quantum chromodynamics (QCD), color superconductivity (CSC) represents a crucial area of study with potential connections to observable physics. This phenomena involves the condensation of quarks into Cooper pairs, known as diquarks, similar to how electrons condense in metallic superconductors. However, what sets QCD apart from electromagnetism is the attractive nature of the strong interaction between quarks, which allows for more direct pairing. The condensation of these quark pairs carries a net color charge, leading to the spontaneous breaking of the $SU(3)$ gauge symmetry of QCD. Note that, this process grants mass to gluons through a Higgs mechanism that, while conceptually related, differs from the Higgs mechanism in the Standard Model (SM). This phenomenon is the basis for QCD CSC \cite{Nam2021}. Moreover, this phase is believed to occur in the inner cores of heavy neutron stars \cite{fadafan2018,Fadafan2021}.

One common way to investigate the QCD CSC phase is through a holographic modeling via the AdS/CFT correspondence \cite{Maldacena1997,Witten98,Gubser1998}. This framework relates a weakly coupled gravity theory in $d$-dimensional anti-de Sitter (AdS) spacetime to a strongly coupled conformal field theory (CFT) on the $(d-1)$-dimensional boundary of that spacetime. Within the context, the CSC phase arises at high chemical potentials and low temperatures which is below the QCD scale. To accommodate this, we introduce an additional compact extra dimension on the boundary that corresponds to the QCD scale. This technique of geometrizing a physical effect has also been employed in classical physics \cite{phan2021curious}. As a result, for the AdS theory to be dual to our four dimensional spacetime universe, the AdS space boundary becomes $R^3\times S^1$, with its bulk spacetime having six dimensions \cite{Basu.et.al.2011}. This contrasts with the five dimensions typically used in holographic QCD \cite{Vu2020} and the four dimensions found in holographic models of metallic superconductivity \cite{Horowitz2008,Hartnoll2008}.

Previous study  \cite{Basu.et.al.2011} has considered the Einstein-Maxwell gravity and the standard Maxwell interaction (i.e. $\mathcal{L}_{\text{Maxwell}}=-\frac{1}{4}F^2$) in six-dimensional AdS spacetime. It is important to note that $\mathcal{L}_{\text{Maxwell}}$ differs from the Maxwell power-law holographic model studied in Cao \cite{Nam2022}. There, an AdS soliton with scalar hair corresponds to the confinement phase, while a Reissner-Nordström (RN) AdS black hole with scalar hair is dual to the deconfinement phase. The scalar hair represents diquarks as s-wave in the CSC phase, which only manifests when the chemical potential $\mu$ exceeds a critical value $\mu_c$ and the temperature $T$ is below a certain threshold $T_c$ \cite{Basu.et.al.2011}. It was believed that color superconductivity could not occur in the confinement phase and only appeared in the deconfinement phase, until Kazuo et al. \cite{Kazuo2019} has demonstrated that, under Einstein-Maxwell gravity, the CSC phase is possible, but limited to cases with a single color $N_c = 1$. In more detail, the Breitenlohner-Freedman (BF) bound \cite{BF1,BF2} for the stability of the scalar field (representing diquark Cooper pair) is broken when $N_c < 1.89$, thus we cannot study the CSC phase with $N_c = 2$. This issue can be addressed by modifying either the gravity framework or the Maxwell law, as suggested in Cao \cite{Nam2021} and Cao \cite{Nam2022}.

Here, we explore the possibility of describing the CSC phase in an $SU(N_c)$ gauge theory with an arbitrary number of colors $N_c$, using a dual quantum gravity theory in a general spacetime dimension, without the need for any modifications. Specifically, we search for the spacetime dimensionality $d$ in which a CSC phase can emerge for $N_c \geq 2$, as the BF bound condition depends non-trivially on the value of $d$ \cite{BF1,BF2}. We start by constructing a general holographic model using Einstein-Maxwell gravity and the standard Maxwell interaction in $d$-dimensional spacetime, as a framework for holographic superconductivity \cite{Emparan2014}. From this rather simple well-established model, we determine the dimensionality $d$ that allows for the emergence of the CSC phase with multiple colors $N_c \geq 2$. The organization of this paper is as follows. In Section \ref{sec2}, we introduce the $d$-dimensional gravitational dual model of interests for the CSC phase transition. In Section \ref{sec3}, we analyze the CSC phase, deriving the conditions on $(d,N_c)$ for the formation of Cooper pairs. Finally, in Section \ref{sec4}, we conclude with the main results and mention some open questions and interesting future directions.

\section{Holographic Model setup \label{sec2}}

We begin with a $d$-dimensional Einstein-Maxwell gravity model to study color superconductivity with gauge symmetry $SU(N_c)$ in the dual quantum field theory. We assume in this field theory, for any value of $N_c$, the confinement phase does not exist, as we will not address confined gauge theories in this work. This assumption, while contradicting QCD observed in our universe at low-energy scale, is rather expected in general e.g. supersymmetric theories $\mathcal{N}=4$, thus does not compromise the generality of our approach. The action for the $d$-dimensional Einstein-Maxwell gravity during the CSC phase transition is given by \cite{Emparan2014}:
 \begin{equation}
 \begin{split}
S=&\int d^dx\sqrt{-g}\Bigg[\mathcal{R}+\frac{(d-1)(d-2)}{L^2}\\
&-\frac{1}{4}F^2-|(\partial_{\mu}-iqA_{\mu})\psi|^2-m^2|\psi|^2 \Bigg] \ ,
 \label{EM_action}
 \end{split}
 \end{equation}
where $F_{\mu\nu}=\partial_{\mu}A_{\nu}-\partial_{\nu}A_{\mu}$ and the cosmological constant is determined by $\Lambda=-\frac{(d-1)(d-2)}{2L^2}$. We then set the AdS radius $L=1$ for convenience. Here, the U(1) gauge field $A_{\mu}$ serves as the dual description analogous to the baryon number current in the CSC phase of QCD or the electric current in metallic superconductivity. The complex scalar field $\psi$ is dual to the boundary Cooper pair scalar field operator; specifically, in the holographic model of the QCD color superconductivity, it corresponds to the diquark Cooper pair scalar field operator (the s-wave CSC). The charge $q$ of this scalar field $\psi$ is associated with the baryon number of the diquark in QCD color superconductivity, and its value is given by 
\begin{equation}
    q=\frac{2}{N_c} \ ,
\label{get_q}
\end{equation}
in which $N_c$ counts number of colors. 

Varying the action \eqref{EM_action} with respect to the vector and scalar fields we obtain the general equation of motion as
\begin{equation}
\label{general eom}
    \begin{split}
        \nabla_{\mu}F^{\mu\nu}-iq[\psi^*(\nabla^{\nu}-iqA^{\nu})\psi-\psi(\nabla^{\nu}+iqA^{\nu})\psi^*]&=0\\
        (\nabla_{\nu}-iqA_{\nu})(\nabla^{\nu}+iqA^{\nu})\psi-m^2\psi&=0
    \end{split}
\end{equation}

To further simplify our model from Eq. \eqref{EM_action}, we focus on s-wave CSC (may be the p-wave or d-wave CSC phase exist but we don't study these in this project), in which the vector field $A_\mu$ and complex scalar field $\psi$ follow the ansatz:
 \begin{equation}
     A_{\mu}dx^{\mu}=\phi(r)dt \ , \  \psi=\psi(r) \ ,
\label{ansatz}
 \end{equation}
where the variations are purely radial. The CSC phase emerges from the condensation of the scalar field Cooper pairs, corresponding to the spontaneous breaking of $U(1)$ symmetry. Assuming that the charge is held fixed, the condensation of the scalar field $\psi$ is driven by the chemical potential, analogous to the baryon chemical potential of quarks in QCD color superconductivity. Near the critical chemical potential, the value of the bulk scalar field approaches zero, and under these conditions, the back reaction of the bulk scalar field on the spacetime metric is negligible. Thus, the back reaction of the matter field is primarily contributed by the $U(1)$ gauge field $A_\mu$ alone.

As mentioned above, this model does not include a confinement phase, and since the CSC phase transition occurs at a critical temperature $T_c$ \cite{Basu.et.al.2011} ,this temperature is associated with the critical chemical potential $\mu_c$ (similar to the QCD CSC phase in the deconfinement phase). Consequently, the spacetime geometry dual to this phase is described by the Reissner-Nordström (RN) planar black hole solution, with the metric given by the following ansatz:
 \begin{equation}
  \label{black hole solution}
 ds^2=r^2 \Big[-f(r)dt^2+h_{ij}dx^idx^j \Big]+\frac{dr^2}{r^2f(r)} \ ,
 \end{equation}
 where $h_{ij}dx^idx^j=dx_1^2+...+dx_{d-2}^2$ is the line element of the $(d-2)$-dimension hypersurface. The event horizon radius $r_+$ satisfies 
 \begin{equation}
      f(r_+)=0 \ .
\label{black_hor}
 \end{equation}
 In the holographic dictionary, the temperature of the boundary field theory is associated with the Hawking temperature of this $d$-dimensional RN planar AdS black hole, i.e. 
 \begin{equation}
 T=T_H\equiv\frac{r_+^2f'(r_+)}{4\pi} \ .
 \label{Hawking}
 \end{equation}

From the Eq.\eqref{general eom}, using the ansatz Eq. \eqref{ansatz}, and Eq. \eqref{black hole solution} we can obtain the classical equations of motion for the temporal component of the vector field $\phi$ and the complex scalar field $\psi$ to be:
  \begin{equation}
  \label{eom}
\begin{split}
&\phi''(r)+\frac{d-2}{r}\phi'(r)-\frac{2q^2\psi^2(r)}{r^2f(r)}\phi(r)=0 \ , 
\\
&\psi''(r)+\left[\frac{f'(r)}{f(r)}+\frac{d}{r}\right]\psi'(r)\\
&+\frac{1}{r^2f(r)}\left[\frac{q^2\phi^2(r)}{r^2f(r)}-m^2\right]\psi(r)=0 \ ,
\end{split}
\end{equation}
in which the blackening function $f(r)$ is given by \cite{Nam2021},\cite{Nam2019}:
 \begin{equation}
 f(r)=1-\left(1+\frac{3\mu^2}{8r_+^2}\right)\left(\frac{r_+}{r}\right)^{d-1}+\frac{3\mu^2r_+^d}{8r^{d+2}}
 \label{blackening}
 \end{equation}
 For a sanity check, in the case of $d=6$-dimensional spacetime, this expression becomes:
 $$f(r)=1-\left(1+\frac{3\mu^2}{8r^2_+}\right)\left(\frac{r_+}{r}\right)^5+\frac{3\mu^2r_+^6}{8r^8} \ , $$ 
 which is equivalent to the holographic model for the QCD CSC in deconfinement region \cite{Kazuo2019}. 

The temperature $T$ of the system lived on the boundary is dual to the Hawking temperature $T_H$ observed in the bulk, as mentioned in Eq. \eqref{Hawking}. Hence:
 \begin{equation}
 T=\frac{r_+^2f'(r_+)}{4\pi}=\frac{1}{4\pi}\left[(d-1)r_+-\frac{9\mu^2}{8r_+}\right] \ .
 \end{equation}
 From the physical condition that these thermal temperatures cannot be negative, we have the constraint for the chemical potential $\mu$:
  \begin{equation}
 \frac{\mu^2}{r_+^2}\leq \frac{8(d-1)}{9} \ .
 \label{chempot_constraint}
 \end{equation}
 
 Near the boundary $(r\rightarrow\infty)$, from the equations of motion Eq. \eqref{eom}, we have the asymptotic forms for the matter fields:
\begin{equation}\label{asymptotic form}
\begin{split}
\phi(r)&= \mu -\frac{\rho}{r^{d-3}} \ , 
\\
\psi(r)&=\frac{J_C}{r^{\Delta_-}}+\frac{C}{r^{\Delta_+}} \ ,
\end{split}
\end{equation}
where $\mu,\rho,J_C$, and $C$ are regarded as the chemical potential, charge density, source, and the condensates vacuum expected value (VEV) of the Cooper pair operator dual to $\psi$ (like the diquark Cooper pair in QCD), respectively. In this case, the conformal dimensions $\Delta_{\pm}$ read:
\begin{equation}
\Delta_{\pm}=\frac{1}{2}\left[(d-1)\pm\sqrt{(d-1)^2+4m^2}\right] \ ,
\end{equation}
and the BF bound -- as follows from \cite{BF1,BF2} -- is:
\begin{equation}
\label{BF bound}
    m^2\geq-\frac{(d-1)^2}{4} \ .
\end{equation}
This is the stability condition for the field $\psi$. To proceed further, we consider setting $\Delta_-=1$ for simplification \cite{Nam2019}, which leads to:
\begin{equation}
    m^2=2-d  \ ,
\label{mass_sq}
\end{equation} 
and therefore obtain $\Delta_+=d-2$. Note that, in this setting, $\psi$ is normalizable and $m^2$ always satisfies Eq. \eqref{BF bound}. Thus, Eq. \eqref{asymptotic form} becomes:
\begin{equation}
    \psi(r)=\frac{J_C}{r}+\frac{C}{r^{d-2}} \ .
\label{psi_func}
\end{equation}


Near the event horizon, the blackening function, the scalar field, and the $U(1)$ gauge field should remain regular. Specifically, the blackening function $f(r)$ vanishes at the event horizon $r=r_+$, as given in Eq. \eqref{black_hor}. Additionally, the matter fields must satisfy the regularity condition at the event horizon, as noted in \cite{Nam2021, Kazuo2019}:
\begin{equation}
    \begin{split}
 &\phi(r_+)=0 \ , \\
        &\psi(r_+)=r_+^2\frac{f'(r_+)\psi'(r_+)}{m^2} \ .
    \end{split}
\label{horizon_condition}
\end{equation}

\section{The emergence of CSC phase with multiple colors \label{sec3}}

When the chemical potential $\mu$ exceeds the critical value $\mu_c$, the Cooper pair condensation occurs. Near that threshold (when $\mu > \mu_c$ but still close to $\mu_c$), $\psi$ should be approximately $0$, so its backreaction can be neglected. Therefore, the bulk configuration is approximately determined by:
 \begin{equation}
 S=\int d^dx\sqrt{-g}\left(\mathcal{R}-2\Lambda-\frac{1}{4}F^2\right) \ .
 \end{equation}
The solution of the gauge field in this case is simply:
\begin{equation}
    \phi(r)=\mu\left[1-\left(\frac{r_+}{r}\right)^{d-3}\right] \ ,
\label{simple_sol}
\end{equation}
which is consistent with Eq. \eqref{asymptotic form} and Eq. \eqref{horizon_condition}. Since the CSC phase involves condensation, we first need a destabilized scalar field. After the collapse of this unstable scalar field, the Cooper pair condensate forms. From the equation of motion for the scalar field, specifically the second equation of \eqref{eom}, we can introduce the following radial effective mass:
 \begin{equation}
    m^2_{\text{eff}}(r) =m^2-\Delta m^2= m^2-\frac{q^2\phi^2(r)}{r^2f(r)} \ .
    \label{mass_eff}
 \end{equation}
With this effective mass, we rewrite the equation of motion Eq. \eqref{eom} of the scalar field:
\begin{equation}
    (\square_{\text{RN}}-m^2_{\text{eff}})\psi_{\text{eff}}=0 \ ,
\end{equation}
where $\square_{\text{RN}}$ is the Laplacian operator defined on the RN planar black hole spacetime background, and the $\psi_{\text{eff}}$ is the effective bulk scalar field. For the Cooper pair condensed state to appear, this field must be unstable somewhere, which is determined from its effective mass. This corresponds to the breaking of the BF bound \cite{BF1,BF2}, leading to the condition:
\begin{equation}
    m^2_{\text{eff}}<\frac{-(d-1)^2}{4} \ .
\label{BF_instab}
\end{equation}
In other words, this is the instability condition for the effective field $\psi_{\text{eff}}$. It is important to note that, while the ``true'' field $\psi$ obey the BF bound, its effective field $\psi_{\text{eff}}$ must break the BF bound for CSC phase to emerge. 

\begin{figure*}[!htbp]
\includegraphics[width=\textwidth]{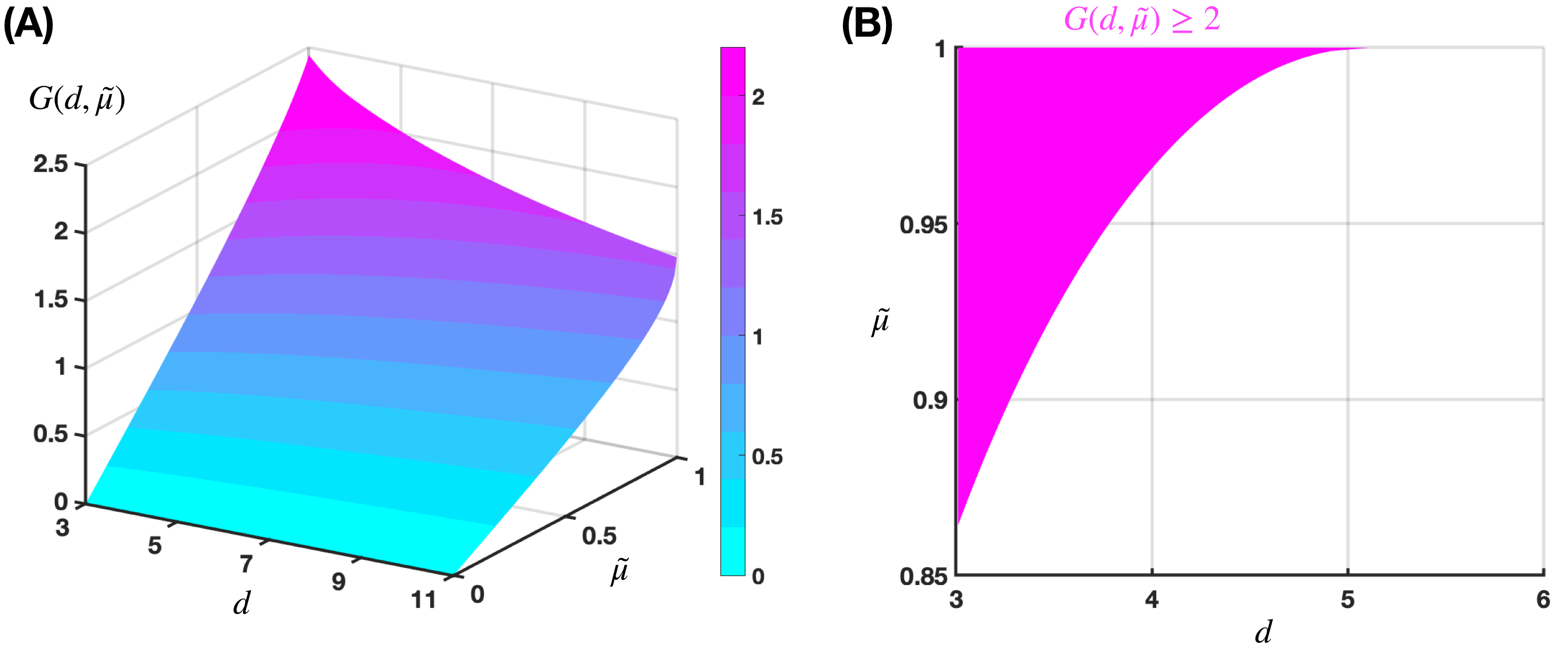}%
\caption{\textbf{Our numerical investigation for $G(d,\tilde{\mu})$ in Eq. \eqref{G_man}.} This calculation was done using MatLab R2023a \cite{MATLAB}. \textbf{(A)} The surface function $G(d,\tilde{\mu})$ inside the region of interests i.e. $(d,\tilde{\mu}) 
\in (3,11] \times [0,1]$. \textbf{(B)} We zoom into the small corner where $G(d,\tilde{\mu})>2$ can be realized.}
\label{fig01}
\end{figure*}

From the condition Eq. \eqref{BF_instab}, we use Eq. \eqref{mass_eff}, in which $m^2=2-d$ as chosen in Eq. \eqref{mass_sq}, to arrive at:
\begin{equation}
\label{BFbound broken}
2-d-\frac{q^2\phi^2(r)}{r^2f(r)}<-\frac{(d-1)^2}{4} \ .
\end{equation}
 This inequality can be solved:
\begin{equation}
\frac{q^2\phi^2(r)}{r^2f(r)}>\frac{(d-3)^2}{4} \ .
\end{equation}
Plug in this expression the values of $\phi(r)$ in Eq. \eqref{simple_sol} and $f(r)$ in Eq. \eqref{blackening}, we obtain:
\begin{equation}\label{conditionofNc}
q^2F(d,\hat{\mu},z) 
>\frac{(d-3)^2}{4} \ ,  
\end{equation}
in which we define the function $F$ to be:
\begin{equation}
F(d,\hat{\mu},z) \equiv \frac{\hat{\mu}^2z^2\left(1-z^{d-3}\right)^2}{1-\left(1+\frac{3\hat{\mu}^2}{8}\right)z^{d-1}+\frac{3\hat{\mu}^2z^{d+2}}{8}} \ ,
\end{equation}
and the new variables are $\hat{\mu}=\frac{\mu}{r_+}$ and $z=\frac{r_+}{r}$.

Note that, when $d=3$, Eq. \eqref{simple_sol} yields $\phi(r)=0$, and the inequality in Eq. \eqref{BFbound broken} is barely not satisfied (i.e. both sides are equal to $-1$). Indeed, this represents an ``edge'' case, hence the condition for the spacetime dimensionality required for the CSC phase to emerge is:
\begin{equation}
    d > 3 \ .
\label{low_dim}
\end{equation}
For $d=3$, the BF bound cannot be broken, and we cannot have condensate with any value of $N_c$. 
For $d=2$, a dual $SU(N_c)$ quantum field theory does not exist because the space boundary vanishes.

Substitute Eq. \eqref{get_q} into Eq. \eqref{conditionofNc}, we arrive at:
 \begin{equation}
     \left(\frac{2}{N_c}\right)^2F(d,\hat{\mu},z)<\frac{(d-3)^2}{4} \ .
 \end{equation}
Thus:
\begin{equation}
    N_c<\frac{4\sqrt{F(d,\hat{\mu},z)}}{d-3} \ .
\label{ineq}
\end{equation}
For the CSC phase to emerge, this inequality should hold in some region of space. Therefore, the upper limit for the number of colors $N_c$ can be determined from: 
    \begin{equation}
 N_c<\frac{4\sqrt{F_{max}(d,\hat{\mu})}}{d-3} \ ,
 \end{equation} 
 where:
\begin{equation}
    F_{\max}(d,\hat{\mu}) \equiv \max_{z\in [0,1]} F(d,\hat{\mu},z) \ .
\end{equation}
Let us further change the variable $\hat{\mu}$ to $\tilde{\mu}$ for simpler constraint condition:
\begin{equation}
\hat{\mu} = \tilde{\mu} \left[\frac{\sqrt{8(d-1)}}3\right] \ \ , \ \     0 < \tilde{\mu} \leq 1 \ ,
\end{equation}
as follows from Eq. \eqref{chempot_constraint}. This constraint arises because the temperature of our system equals the Hawking temperature of the AdS black hole, which must be greater than or equal to zero. We then define the following function:
\begin{equation}
    G(d,\tilde{\mu}) \equiv \frac{4 \sqrt{F_{\max}(d,\tilde{\mu}) }}{d-3} \ ,
  \label{G_man}
\end{equation}
in which the condition for color superconductivity is met when the inequality 
\begin{equation}
    G(d,\tilde{\mu})> N_c
    \label{G_cond}
\end{equation}
holds, as follows from Eq. \eqref{ineq}. 

Figure \ref{fig01}A presents our numerical investigation of $G(d,\tilde{\mu})$ for $d \in (3,11]$, covering the dimensionality from the lowest value satisfied Eq. \eqref{low_dim} to the highest-dimensional supergravity \cite{Natsuume2015} (where fundamental objects are conjectured to be membranes \cite{Vo2024Size} rather than strings \cite{Karliner1988Size}). We find no region where $G(d,\tilde{\mu}) > 3$ and only a small area, shown in Fig. \ref{fig01}B, where $G(d,\tilde{\mu}) > 2$. From the condition Eq. \eqref{G_cond}, our results suggest that color superconductivity for $N_c=2$ can be described via holography using only Einstein-Maxwell gravity, but this is valid only in the case of $d=4$. For $N_c \geq 3$, the Einstein-Maxwell model cannot describe the CSC phase for any number of bulk dimensions $d$. Therefore, it is necessary to employ modified gravity, such as Einstein-Gauss-Bonnet \cite{Nam2021}, or adjust Maxwell's equations \cite{Nam2022} to study color superconductivity for $N_c \geq 2$ with $d>4$.

\ \


\section{Discussion \label{sec4}}

We have found that, for integer $d$-dimensional bulk spacetime, only for when $d=4$, can we construct one simple holographic model for CSC with $N_c=2$, using Einstein-Maxwell gravity and the standard Maxwell interaction. This setup enables the study of the CSC phase with $N_c=2$ in $3$-dimensional spacetimes through holography, but not with $N_c\geq2$ in any other dimensions. Note that, we are not yet working within a confined gauge theory, which would exhibit a confinement-deconfinement phase transition in $d$-dimensions. Further challenges include developing holographic models for e.g. p-wave and d-wave CSC, as well as examining the Josephson junction effect in the CSC phase through holography. These aspects is worth exploring in future works.



Here, we have focused on integer bulk dimensions. An interesting direction is holographic modeling in fractional dimensions, where 
$d$ is continuous. As shown in Fig. \ref{fig01}, this approach would place the dual gauge theory in a spacetime dimension below four, potentially enabling experimental realizations. Recent advances in fractal lattice design \cite{Kempkes2019Design} underscore the promise of this research direction, as fractional-dimensional structures are common in nature \cite{mandelbrot1982fractal}. Investigating such models could be especially relevant for condensed matter and quantum gravity, where these unconventional geometries might reveal novel phenomena, as evidenced in fluid dynamics \cite{phan2024vanishing} and soft matter physics \cite{phan2020bacterial}.

\section{Acknowledgements}

We would like to thank Nam H. Cao for his helpful feedback and insightful discussions. We also acknowledge Trung V. Phan for his assistance in generating Fig. \ref{fig01} and editing this manuscript. Additionally, we are grateful to Dmitry Voskresensky for his valuable discussions on color superconductivity.

\bibliography{main}
\bibliographystyle{unsrt}

\end{document}